\documentclass[a4,preprint,pre,showpacs]{revtex4}

\def\EQ{\begin{equation}}
\def\EN{\end{equation}}
\def\EQA{\begin{eqnarray}}
\def\ENA{\end{eqnarray}}
\usepackage{amsmath}
\usepackage[dvips]{graphicx}
\usepackage{natbib}
\usepackage{bm}

\def\uu{{\bf u}}
\def\UU{{\bf U}}
\def\bb{{\bf b}}
\def\BB{{\bf B}}

\def\VV{{\bf U}}

\def\A{\mathcal{A}}

\def\R{\mathcal{R}}
\def\S{\mathcal{S}}
\def\W{\mathcal{W}}
\def\V{\mathcal{V}}
\def\M{\mathcal{M}}

\def\T{\mathcal{T}}

\begin{document}

\title{Turbulent transport and dynamo in sheared MHD turbulence with a non-uniform magnetic field} 
\author{Nicolas Leprovost and Eun-jin Kim}
\affiliation{Department of Applied Mathematics, University of Sheffield, Sheffield S3 7RH, UK}

\begin{abstract}
We investigate three-dimensional magnetohydrodynamics turbulence in the presence of velocity and magnetic shear (i.e. with both a large-scale shear flow and a non-uniform magnetic field). By assuming a turbulence driven by an external forcing with both helical and non-helical spectra, we investigate the combined effect of these two shears on turbulence  intensity and turbulent transport represented by turbulent diffusivities (turbulent viscosity, $\alpha$ and $\beta$ effect) in Reynolds-averaged equations. We show that turbulent transport (turbulent viscosity and diffusivity) is quenched by a strong flow shear and a strong magnetic field.  For a weak flow shear, we further show that the magnetic shear increases the turbulence intensity while decreasing the turbulent transport. In the presence of a strong flow shear, the effect of the magnetic shear is found to oppose the effect of flow shear (which reduces turbulence due to shear stabilization) by enhancing turbulence and transport, thereby weakening the strong quenching by flow shear stabilization. In the case of a strong magnetic field (compared to flow shear), magnetic shear increases turbulence intensity and quenches turbulent transport.
\end{abstract}


\maketitle

\section{Introduction}
Most geophysical and astrophysical bodies are composed of electrically conducting fluids (liquid iron for the Earth and plasma for stars, interstellar medium, etc.). The evolution of magnetic field ${\bf B}$ and velocity ${\bf U}$ in these systems are often described by a simplified model given by the incompressible magnetohydrodynamics (MHD) equations \citep{Moffatt78}:
\EQA
\label{eqMHD}
\partial_t \VV + \VV \cdot {\bf \nabla} \VV &=& - {\bf \nabla} P + \BB \cdot {\bf \nabla} \BB  + \nu \Delta \VV + {\bf f} \; , \\  \label{eqMHD2}
\partial_t \BB + \VV \cdot {\bf \nabla} \BB &=&  \BB \cdot {\bf \nabla} \VV + \eta \Delta \BB  \; , \\  \label{eqMHD3}
{\bf \nabla} \cdot \VV &=& {\bf \nabla} \cdot \BB = 0 \; ,
\ENA
where ${\bf B}$ is the Alfv\'en velocity (${\bf B} = {\bf M} / \sqrt{\mu \rho}$ where ${\bf M}$ is the magnetic field measured in Tesla; therefore ${\bf B}$ has the dimension of velocity), $p$ is the total (hydrodynamical + magnetic) pressure and $\nu$ and $\eta$ are the molecular viscosity and diffusivity, respectively. Eq. (\ref{eqMHD}) is the Navier-Stokes equation including the Lorentz force which describes the effect of the magnetic field on the velocity field and an external forcing ${\bf f}$, which is assumed to be at small scales. Eq. (\ref{eqMHD2}) describing the evolution of the magnetic field is called the induction equation and can be derived from Maxwell's equations and Ohm's law.

In most astrophysical objects, velocity and magnetic fields are observed to exist on a broad range of length and time scales. In order to characterize the evolution of fields on these scales, theories, such as mean-field dynamo \cite{Moffatt78,Krause80}, decompose the fields into a mean and fluctuating parts and parameterize the effect of the small-scale (unresolved) fields on the large scale fields in terms of transport coefficients. Specifically, expressing ${\bf B} = \langle {\bf B} \rangle + {\bf b}$  and ${\bf U} = \langle {\bf U} \rangle + {\bf u}$, where the $\langle \bullet \rangle$ stands for an average on the realization of the small-scale fields, substituting this decomposition into Eqs. (\ref{eqMHD}-\ref{eqMHD3}) and averaging yield the following mean-field equations:
\EQA
\label{Induction2}
\partial_t \langle {\bf U} \rangle + \langle {\bf U} \rangle \cdot {\bf \nabla} \langle \VV \rangle &=&  - {\bf \nabla} \langle p \rangle + \langle \BB \rangle \cdot {\bf \nabla} \langle \BB \rangle + \nu \nabla^2 \langle {\bf U} \rangle - {\bf \nabla} \cdot {\bf \mathcal{S}} \; , \\  \label{Induction2bis}
\partial_t \langle {\bf B} \rangle + \langle {\bf U} \rangle \cdot {\bf \nabla} \langle \BB \rangle &=&  \langle \BB \rangle \cdot {\bf \nabla} \, \langle {\bf U} \rangle + \eta \nabla^2 \langle {\bf B} \rangle + {\bf \nabla} \times {\bf \mathcal{E}} \; , \\  
{\bf \nabla} \cdot \langle \VV \rangle &=& {\bf \nabla} \cdot \langle \BB \rangle = 0 \; .
\ENA
The main challenge is to express the stress tensor $\mathcal{S}=\langle \uu \otimes \uu \rangle - \langle \bb \otimes \bb \rangle$ in Eq. (\ref{Induction2}) and the electromotive force ${\bf \mathcal{E}} = \langle {\bf u} \times {\bf b} \rangle$ in Eq. (\ref{Induction2bis}) in terms of the large-scale variables $\langle {\bf U} \rangle$ and $\langle {\bf B} \rangle$. In the absence of non-diffusive fluxes, the stress can be expressed as $\mathcal{S}_{ij} = - \mathcal{N}^T_{ijkl} \partial_k \langle U_l \rangle$ where $\mathcal{N}^T_{ijkl}$ is called the turbulent viscosity tensor, which can add to the molecular viscosity in Eq. (\ref{Induction2}). On the other hand, the electromotive force is usually assumed to depend linearly on the mean magnetic field and only the two first term are kept (proportional to the magnetic field and its first derivative). Following \citet{Radler06}, this expansion can be written as:
\EQ
\label{ElectromotiveBis}
\mathcal{E}_i = \alpha_{ij} \langle B_j \rangle + [{\bf \mu \times B}]_i - \beta_{ij} ({\bf \nabla \times} \langle {\bf B} \rangle)_j  - [{\bf \delta \times} ({\bf \nabla \times} \langle {\bf B} \rangle)]_i - \zeta_{ijk} [\nabla_j \langle B_k \rangle + \nabla_k \langle B_j \rangle]/2 \; .
\EN
The first term on the right-hand side (RHS) of Eq. (\ref{ElectromotiveBis}) is the $\alpha$ effect which can be shown to generate magnetic field on large scale for a helical turbulence. It is thus a perfect candidate to explain magnetic fields in systems influenced by Coriolis force such as in the stellar convection zones. The second term on the RHS describes a transport of magnetic flux by turbulence. The third and fourth term in Eq. (\ref{ElectromotiveBis}) can be described by introducing an anisotropic turbulent diffusivity. The last term proportional to $\kappa$ does not allow a simple interpretation. The presence of the additional terms besides the $\alpha$ and $\beta$ effect is possible only for anisotropic or/and inhomogeneous turbulence.

\begin{figure}
\begin{center}
\includegraphics[scale=0.7,clip]{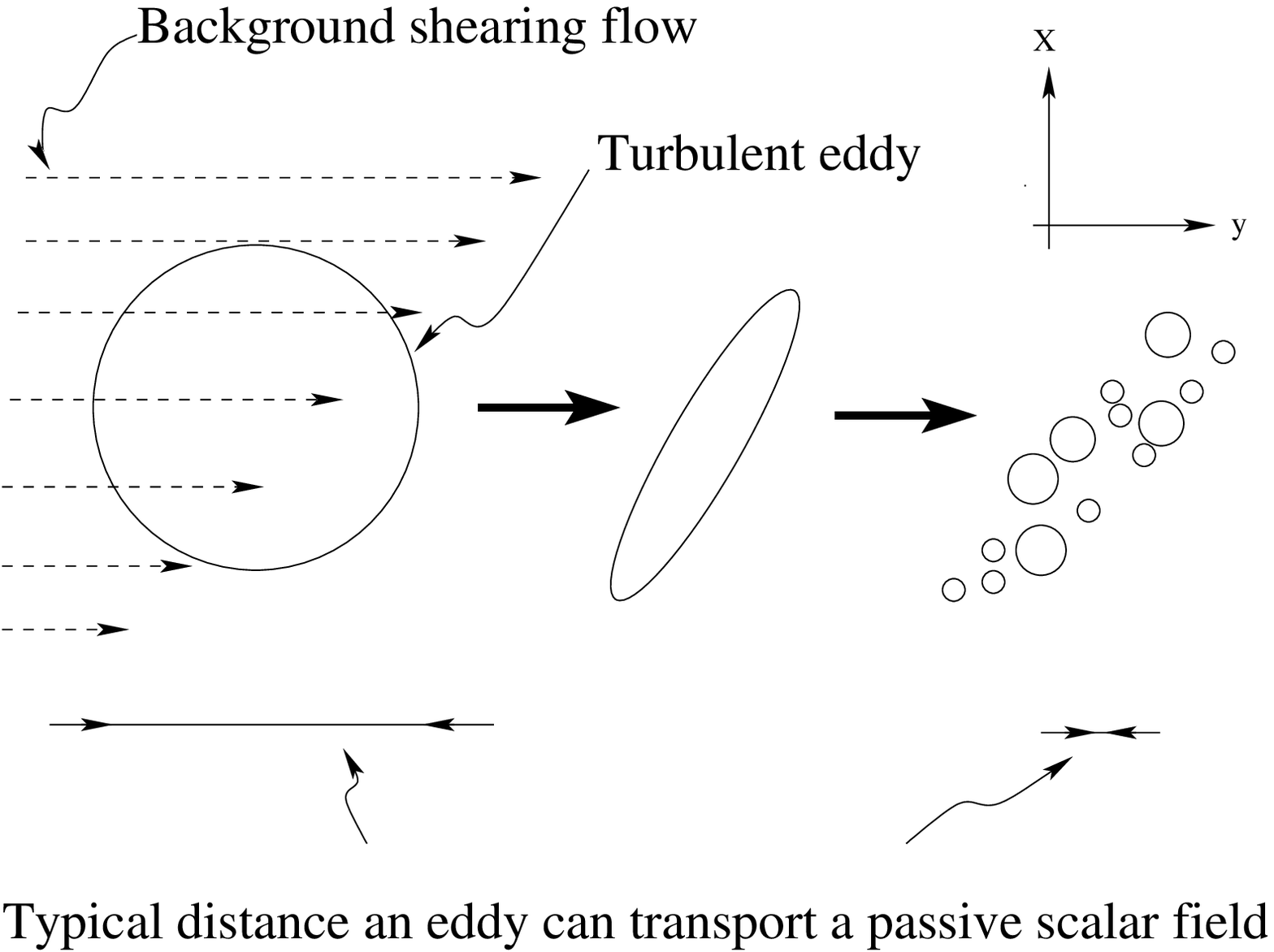}
\caption{\label{ShearEff} Sketch of the effect of shear on a turbulent eddy.}
\end{center}
\end{figure}

There has been accumulating evidence that a strong shear reduces turbulent transport via shear stabilization by flow shear \cite{Burrell97}. This is basically because shear advects turbulent eddies differentially, elongating and distorting their shapes, thereby rapidly generating small scales which are ultimately disrupted by molecular dissipation on small scales (see Fig. \ref{ShearEff}). That is, flow shear facilitates the cascade of various quantities such as energy to small scales (i.e. direct cascade) in the system, enhancing their dissipation rate. As a result, turbulence level as well as turbulent transport of these quantities can be significantly reduced compared to the case without shear. Another important consequence of shearing is to induce anisotropic transport and turbulent level since flow shear directly influences the component parallel to itself (i.e. the $x$ component in Fig. 1) via elongation while only indirectly the other two components (i.e. the $y$ and $z$ components in Fig. 1) through enhanced dissipation and incompressibility. Indeed, the flow shear has been shown to significantly reduce the turbulence intensity and the turbulent transport of angular momentum, particle mixing and magnetic diffusion both in hydrodynamics \cite{Kim05,2Shears} and magnetohydrodynamics \cite{Kim06}. In \cite{Quenching}, by assuming a uniform large-scale magnetic field, we showed that the $\alpha$ effect is quenched by flow shear and magnetic field.  

In this paper, we first examine the effect of non-uniform magnetic field on the linear stability of MHD fluids. We then proceed to investigate the effect of a non-uniform magnetic field on MHD turbulence. A consistent derivation of turbulent magnetic diffusion ($\beta$ effect) of non-uniform field in 3D MHD is crucial since there has been controversy over the $\beta$ suppression by strong magnetic field \cite{Gruzinov95}. In comparison, in 2D MHD, the $\beta$ effect was shown to be severely quenched by magnetic field and also by shear flow \cite{Kim01}. Furthermore, in the presence of shear flow, another interesting effect of magnetic shear in non-uniform magnetic field is that the latter can interfere with flow shear and thus weaken the quenching of turbulence by flow shear, as shown in 3D reduced MHD turbulence \cite{Diamond05,Kim07}. We investigate if this result holds in 3D MHD by quantifying the effect of magnetic shear on turbulence intensity and turbulent transport. The reminder of the paper is organized as follows. In Section \ref{Model}, we present the main governing equations and the quasi-linear approximation that is used to solve for the turbulent fields. We then present the calculations of the transport coefficients (turbulent viscosity and $\alpha$ effect) in the limit where the uniform component of the large-scale magnetic field is negligible in Sec. \ref{WeakMagnetic} and then in the other extreme limit where it is very strong (compared to shear flow)  in Sec. \ref{StrongMagnetic}. Conclusions are provided in Sec. \ref{Discussion}.

\section{Model}
\label{Model}
\subsection{Magnetohydrodynamical equations}
We use the similarity of the Navier-Stokes (\ref{eqMHD}) and induction equation (\ref{eqMHD2}) and work in terms of Elsasser variables \citep{Elsasser50} instead of the (fluctuating) velocity and magnetic fields: ${\bf \Psi}^+ = \VV + \BB$  and ${\bf \Psi}^- = \VV - \BB$. Assuming a unit magnetic Prandtl number ($\nu = \eta$), the MHD equations (\ref{eqMHD}-\ref{eqMHD3}) can be rewritten as:
\EQA
\partial_t {\bf \Psi}^+ + {\bf \Psi}^- \cdot {\bf \nabla} {\bf \Psi}^+  &=&  - {\bf \nabla} P + \nu \Delta {\bf \Psi}^+ + {\bf f}   \; , \\  
\partial_t {\bf \Psi}^- + {\bf \Psi}^+  \cdot {\bf \nabla} {\bf \Psi}^-  &=&  - {\bf \nabla} P + \nu \Delta {\bf \Psi}^- + {\bf f}   \; , \\  
{\bf \nabla} \cdot {\bf \Psi}^+ &=& {\bf \nabla} \cdot {\bf \Psi}^- = 0 \; .
\ENA
To study the effect of shear flows and magnetic fields on small-scale turbulence, we prescribe a large scale flow of the form $\langle \UU \rangle = - x \A {\bf e_y}$ and a non-uniform large-scale magnetic field $\langle {\bf B} \rangle = (B_0 - B_1 x) \, {\bf e_y}$. The chosen configuration (with parallel velocity and magnetic field) ensures that there is no direct influence of $\langle \UU \rangle$ on $\langle {\bf B} \rangle$. To solve the equations for the Elsasser variables, ${\bm \psi}^+ = {\bf \Psi}^+ - \langle {\bf \Psi}^+ \rangle = {\bf \Psi}^+ - \langle \UU + \BB \rangle$ and ${\bm \psi}^- = {\bf \Psi}^- - \langle {\bf \Psi}^- \rangle = {\bf \Psi}^+ - \langle \UU - \BB \rangle$, we use the quasi-linear approximation assuming that the interaction between fluctuating fields is negligible compared to the interaction between large and small-scale fields. The equations for the fluctuating fields can then be written as:
\EQA
\label{eqRDT}
\partial_t {\bm \psi}^+ + \langle {\bm \Psi}^- \rangle \cdot {\bm \nabla} {\bm \psi}^+ + {\bm \psi}^- \cdot {\bm \nabla} \langle {\bm \Psi}^+ \rangle &=&  - {\bm \nabla} p + \nu \Delta {\bm \psi}^+ + {\bm f}   \; , \\  \label{eqRDT2}
\partial_t {\bm \psi}^- + \langle {\bm \Psi}^+ \rangle \cdot {\bm \nabla} {\bm \psi}^- + {\bm \psi}^+ \cdot {\bm \nabla} \langle {\bm \Psi}^- \rangle &=&  - {\bm \nabla} p + \nu \Delta {\bm \psi}^- + {\bm f}   \; , \\  \label{eqRDT3}
{\bm \nabla} \cdot {\bm \psi}^+ &=& {\bm \nabla} \cdot {\bm \psi}^- = 0 \; ,
\ENA
where $p$ is the fluctuation in the pressure. To solve these equations, we use vanishing initial conditions: ${\bm \psi}^+ = {\bm \psi}^- = 0$ at $t=t_0$. An equilibrium is then reached at  long times when the power injected by the forcing balances the dissipation. To solve Eqs. (\ref{eqRDT}-\ref{eqRDT3}), we introduce a time-dependent Fourier transform:
\EQ
Y({\bf x},t) = \frac{1}{(2\pi)^3} \int d^3 {\bf k} e^{i\bigl[{ k_x(t)} x + k_y y + k_z z\bigr]}
\tilde{Y}({\bf k},t) \; ,
\EN
where we choose $k_x(t) = k_x(t_0)  + \A k_y (t-t_0)$ to account non-perturbatively  for the effect of the non-uniform components of $\langle \UU \rangle$ and $\langle \BB \rangle$.  Fourier-transformation of Eqs. (\ref{eqRDT}-\ref{eqRDT3}) leads to the following equations:
\EQA
\label{System}
\partial_t \tilde{\psi}^{\pm}_i \mp k_y B_1 \partial_{k_x} \tilde{\psi}^{\pm}_i - (\A \pm B_1) \, \tilde{\psi}^{\mp}_i \delta_{i2} &=& \pm i B_0 k_y \tilde{\psi}^{\pm}_i - i k_i\tilde{p} - \nu k^2  \tilde{\psi}^{\pm}_i + \tilde{f}_i \; , \\ \label{SystemBis}
k_x \tilde{\psi}_x^{\pm} + k_y \tilde{\psi}_y^{\pm} + k_z \tilde{\psi}_z^{\pm} &=& 0 \; .
\ENA
Note that the time-dependence of the wave number cancels exactly the advection by the mean velocity shear (see \citet{Kim05} for details) and that the second term on the left hand side of Eq. (\ref{System}) is obtained by Fourier transform of the advection by the mean magnetic shear in the following way:
\EQA
\mathcal{FT}(B_1 x \partial_y \psi^{\pm}_i) &=& \int d^3 x \; e^{- i\bigl[{ k_x(t)} x + k_y y + k_z z\bigr]} B_1 x \partial_y \psi^{\pm}_i \\ \nonumber
&=& i k_y B_1 \int d^3 x \; i \partial_{k_x} \left(e^{- i\bigl[{ k_x(t)} x + k_y y + k_z z\bigr]}\right)  \psi^{\pm}_i \\ \nonumber
&=& - k_y B_1 \partial_{k_x} \tilde{\psi}^{\pm}_i \; .
\ENA
Changing the time variable from $t$ to $\tau = k_x(t) / k_y = k_x(t_0) / k_y + \A (t-t_0)$, the first two terms on the left-hand side of Eq. (\ref{System}) can be grouped together as $\partial_t = \A \partial_\tau$ and $k_y \partial_{k_x} = \partial_\tau$. Assuming the forcing to be incompressible, Eqs. (\ref{System}-\ref{SystemBis}) can be rewritten:
\EQA
\label{System2bis}
(1\mp\R) \partial_\tau \tilde{\psi}^{\pm}_i - (1\pm\R) \tilde{\psi_x}^{\mp} \delta_{i2} &=& \pm i \gamma \tilde{\psi}^{\pm}_i \\  \nonumber 
 - \frac{1}{g^2+\tau^2} \left(
\begin{array}{c}
\tau \\
1 \\
b
\end{array}
\right) \left[(1-\R) \tilde{\psi}^{+}_x + (1+\R) \tilde{\psi}^{-}_x   \right]  &-& \xi (g^2+\tau^2) \tilde{\psi}^{\pm}_i + \frac{1}{\A} \tilde{f}_i \; , \\ \label{System2}
\tau \tilde{\psi}_x^{\pm} + \tilde{\psi}_y^{\pm} + b\tilde{\psi}_z^{\pm} &=& 0 \; .
\ENA
Here, $\R = B_1/ \A$ is the ratio of the magnetic shear to the velocity shear; $\gamma=B_0 k_y/ \A$ is the ratio of the Alfv\'en frequency to the flow shear;  $b= k_z / k_y$ and $g^2=1+b^2$; $\xi=\nu k_y^2 / \A$. Note that Eqs. (\ref{System2bis}-\ref{System2}) are invariant under the following transformation: $\gamma \leftrightarrow - \gamma$, $\R \leftrightarrow -\R$ and $\tilde{\bm \psi}^+ \leftrightarrow \tilde{\bm \psi}^-$. Consequently, $\tilde{\bm \psi}^{-}$ can be obtained from $\tilde{\bm \psi}^{+}$ (and vice versus) by changing the sign of $\gamma$ and $\R$. Using the following variables:
\EQ
\label{ChgVar}
\tilde{\psi}_x^+(\tau) = \frac{\phi^+(\tau)}{(1-\R) (g^2+\tau^2)} \quad \text{and} \quad \tilde{\psi}_x^-(\tau) = \frac{\phi^-(\tau)}{(1+\R) (g^2+\tau^2)} \; ,
\EN
the $x$-component of Eq. (\ref{System2bis}) can be rewritten:
\EQ
\label{NewSyst}
\partial_\tau \phi^+ = \frac{i \gamma \phi^+}{1-R} + \frac{\tau}{g^2+\tau^2} \left[\phi^+ - \phi^-\right] - \frac{\xi}{1-\R} (g^2+\tau^2) \phi^+ + \frac{(g^2+\tau^2)}{\A} \tilde{f}_x\; . \\  
\EN
The coupled equations for $\phi^+$ and $\phi^-$ in Eq. (\ref{NewSyst}) can be combined to a closed equation for $\phi^+$ :
\EQA
\label{OneEquation}
\partial_\tau^2 \phi^{+} &+& \left[-\frac{1}{\tau}  + \frac{2 \xi (g^2+\tau^2)}{1-\R^2} - \frac{2 i \gamma \R}{1-\R^2} \right] \partial_\tau \phi^{+} \\  \nonumber
&+& \left[\frac{i \gamma}{(g^2+\tau^2)} \left(\frac{\gamma}{(1-\R)\tau} -\frac{\tau}{1+\R}  \right)  + \frac{\xi}{1-\R} \left( \frac{\tau^2-g^2}{\tau} + \frac{2 \R \tau}{1+\R}\right) + \frac{\xi^2 (g^2+\tau^2)^2 +\gamma^2}{1-\R^2} \right] \phi^{+} \\  \nonumber
&=& \tau \partial_\tau\left[\frac{(g^2+\tau^2) \tilde{f}_x}{\A \tau} \right] + \frac{i\gamma + \xi (g^2+\tau^2) }{1+\R} (g^2+\tau^2) \frac{\tilde{f}_x}{\A} \; .
\ENA
Two initial conditions are needed in order to solve Eq. (\ref{OneEquation}). At $\tau = \tau_0 = k_x(t_0) / k_y$, we assume that initially there is no velocity and magnetic perturbations ($\phi^{+}(\tau=\tau_0) = \phi^{-}(\tau=\tau_0) = 0$). The use of $\phi^{\pm}(\tau_0)=0$ in  Eq. (\ref{NewSyst}) gives us  $[\partial_\tau \phi^{+}] (\tau=\tau_0) = \tilde{f}_x(\tau=\tau_0) (g^2+\tau_0^2) / \A$ as the second initial condition. As Eq. (\ref{OneEquation}) cannot be solved in general case, we will consider the two cases of weak and strong magnetic field, given by $\gamma=B_0 k_y /\A \ll 1$ and $\gamma \gg 1$ respectively. Note that the weak magnetic field limit does not restrict the magnitude of the magnetic shear. The only constraint on the magnetic shear comes from our assumption that the system is stable (i.e. $\vert \R \vert < 1$ as shown in the next subsection).

\subsection{linear Stability}
\label{LinearStab}
We start by studying the stability of the large scale fields $\langle \UU \rangle$ and $\langle \BB \rangle$ by considering the behavior of the perturbations of Eq. (\ref{OneEquation}) in the long time limit. As Eq. (\ref{OneEquation}) is a second order differential equation in $\tau$, the homogeneous equation has two independent solutions $\phi_1(\tau)$ and $\phi_2(\tau)$. Using WKB theory (see Appendix \ref{WKBLargeTau} for details), we can show that these two functions have the following asymptotic behavior in the large $\tau$ limit: 
\EQA
\label{LargeTau}
\phi_1(\tau) &\sim& \frac{1}{\tau^2}
 \exp\left[-\frac{\xi}{1+\R} Q(\tau) - \frac{i \gamma}{1+\R} \tau \right] \; , \\ 
\label{LargeTau2}
\phi_2(\tau) &\sim& \tau \exp\left[-\frac{\xi}{1-\R} Q(\tau) + \frac{i \gamma}{1-\R} \tau \right]  \; ,
\ENA
where $Q(x) = g^2 x + x^3 / 3$ and the $\sim$ symbol stands for asymptotic behavior in the large $\tau$ limit. Eqs. (\ref{LargeTau}-\ref{LargeTau2}) show that if $\vert \R \vert \geq 1$ one of the solutions of the homogeneous problem is exponentially divergent for large $\tau$. This is a similar result as that found by \cite{Chen90} who studied the resistive tearing instability in the context of fusion plasmas. Interestingly, the tearing instability is stabilized by flow shear if $\vert \R \vert < 1$. Note that the limit $\R \rightarrow 0$ is singular as we do not recover the result of \cite{Kim05} with no magnetic shear. This is due to an additional dissipative layer imposed by the magnetic field. Consequently, the effect of the magnetic shear, is two-fold. On the one hand, it renormalizes the diffusion rate $\xi$ in the exponential factor to $\xi / (1-\R)$ and $\xi /(1+\R)$ (with a continuous limit to the case $\R=0$). On the other hand, magnetic shear changes the scaling with $\tau$ in front of the exponential in a non-continuous fashion (note that the limit $\xi=0$ is also singular as this asymptotic behavior does not recover the result for the ideal system that can be computed exactly). In the following, we restrict our study to the case where the system is stable ($\vert \R \vert <1$).

\subsection{Transport coefficients}
\label{TransportCoeff}
Our main interest is in the total stress and the electromotive force, which determine the growth/decay of the large-scale velocity field and the large-scale magnetic field, respectively. The assumption of a large scale flow is in the $y$ direction and depending only on the $x$ coordinate has two implications for the stress. First, only the components $\mathcal{N}^T_{ijxy}$ do not vanish. Second, only the $y$-component of Eq. (\ref{Induction2}) is of interest and the divergence of the Reynolds stress reduces to $\partial_x S_{xy}$ (the two other terms involves derivative with respect to $y$ and $z$). Consequently, in the following, we are interested in only one component of the total stress $S$ defined as:
\EQ
\label{TurbVisc}
\mathcal{S} \equiv \mathcal{S}_{xy} = \langle u_x u_y \rangle - \langle b_x b_y \rangle = \frac{1}{2} \langle \psi_x^+ \psi_y^- + \psi_x^- \psi_y^+ \rangle \; .
\EN
Note that this total stress consists of the difference between the Reynolds stress $\langle u_x u_y \rangle$ and Maxwell stress $\langle b_x b_y \rangle$. In the following, we refer to turbulent viscosity $\nu^T$ as the only component of interest $\nu^T = \mathcal{N}^T_{xyxy}$. For the assumed shear flow $\langle {\bf U} \rangle  = - \A x {\bf e_y}$, the turbulent diffusivity  can be computed as $\mathcal{S} = \nu^T \A$.

Similarly, as we chose the large-scale magnetic field to depend only on $x$, the only components of ${\bf \nabla \times \mathcal{E}}$ are the one in the $y$ and $z$ direction given by:
\EQA
\label{BS2}
\mathcal{E}_y = \langle u_z b_x - u_x b_z \rangle = \langle \psi_x^+ \psi_z^- - \psi_x^- \psi_z^+ \rangle /2 \; , \\  \label{BS2ter}
\mathcal{E}_z = \langle u_x b_y - u_y b_x \rangle = \langle \psi_y^+ \psi_x^- - \psi_y^- \psi_x^+ \rangle /2 \; .
\ENA
For our chosen configuration of the magnetic field which depends only on $x$, the electromotive force has only the following terms (see \cite{Rogachevskii03} for a general expression of the electromotive force for arbitrary shear flows):
\EQA
\mathcal{E}_y &=& \alpha_{yy} B_0 \; , \\  
\mathcal{E}_z &=& (\alpha_{zy} + \mu_z) B_0 + \beta B_1\; ,
\ENA
where $\alpha_{yy}$ and $\alpha_{zy}$ are the components of anisotropic (due to shear flow) $\alpha$ effect, $\mu_z$ is the turbulent transport of magnetic flux and $\beta$ characterizes the turbulent $\beta$ effect. Note here that only these three coefficients are present in our configuration. In particular, phenomena such  as the ${\bf \Omega \times J}$ \cite{Radler06} and shear current effects \cite{Rogachevskii03}, which have been advocated for turbulence affected by rotation and shear, are absent here. For instance the shear current effect vanishes because the large-scale vorticity and the curl of the magnetic field are parallel to each other \cite{Rogachevskii03}.

\subsection{Forcing}
To calculate the correlation functions involved in the transport coefficients [see Eq. (\ref{TurbVisc}-\ref{BS2ter})], we consider an incompressible forcing which is spatially homogeneous and temporally stationary  with a short correlation time $\tau_f$. Specifically, in Fourier space, the correlation function of the forcing is taken as:
\EQ
\label{Forcing}
\langle \tilde{f}_i({\bf k_1},t_1) \tilde{f}_j({\bf k_2},t_2) \rangle = \tau_f \, (2\pi)^3 \delta({\bf k_1}+{\bf k_2}) \, \delta(t_1-t_2)  \kappa_{ij}({\bf k_2}) \; .
\EN
As noted previously, the $\alpha$ effect can be linked to the helicity of the turbulent flow. Consequently, we consider a forcing with both a symmetric part (with energy spectrum $E$) and a helical part (with helicity spectrum $H$) given by:
\EQ
\kappa_{lm}({\bf k}) = E(k) \left(\delta_{lm} - \frac{k_l k_m}{k^2} \right) + i \epsilon_{lmp} k_p \frac{H(k)}{k^2} \; . 
\EN
In the following, the turbulence intensity, turbulent viscosity and $\alpha$ effect are expressed in terms of the kinetic energy $e_0=\langle {\bf u}^2 \rangle$ and helicity $h_0=\langle {\bf u} (\cdot {\bf \nabla \times u})$ of the flow created by the forcing $f$ in absence of shear and magnetic field:
\EQA
\label{NoShear}
e_0 &=& \frac{\tau_f}{(2\pi)^2} \int_0^{+\infty} d k \frac{E(k)}{\nu} \; , \\  \label{NoShear2}
h_0 &=& \frac{\tau_f}{(2\pi)^2} \int_0^{+\infty} d k \frac{H(k)}{\nu}  \; .
\ENA
The presence of the viscosity $\nu$ in the denominator of Eqs. (\ref{NoShear}-\ref{NoShear2}) is due to the fact that without dissipation ($\nu=0$), there is nothing to dissipate the energy injected by the (small-scale) forcing causing the accumulation of small-scale fields. Therefore, the growth of kinetic energy or transport is unbounded. In appendix \ref{AppendixA}, we show the derivation of Eq. (\ref{NoShear}) and Eq. (\ref{NoShear2}).

\section{Weak Magnetic field ($\gamma \ll 1$)}
\label{WeakMagnetic}
In this section, we investigate the influence of magnetic and flow shear on turbulence properties in the limit of a very weak magnetic field $B_0$ (i.e. $\gamma \ll 1$). In the following, we thus keep the magnetic field only to the lowest order to compute the $\alpha$ effect. As it seems impossible to exactly solve Eq. (\ref{OneEquation}) even for $\gamma=0$, we obtain our results in two different limits of weak ($\xi =\nu k_y^2 /\A \gg 1$) and strong ($\xi \ll 1$) shear in comparison with the diffusion rate ($\nu k_y^2$).

\subsection{Weak Flow Shear: $\xi \gg 1$} 
\label{WeakShear}
In the case of a shearing rate much weaker than diffusion rate ($\xi = \nu k_y^2 / \A \gg 1$), we can perform a WKB analysis (very similar to that of appendix \ref{WKBLargeTau}) of Eq. (\ref{OneEquation}) by using $\xi \gg 1$ as a large parameter, obtaining the following two solutions of the homogeneous system (\ref{OneEquation}):
\EQA
\label{WeakHomogeneous}
\phi_1(\tau) &=& \frac{\tau}{(g^2+\tau^2)^{3/2}} \; \exp\left[-\frac{\xi}{1+\R}Q(\tau)-\frac{i \gamma}{1+\R} \tau \right] \equiv  \frac{\tau}{(g^2+\tau^2)^{3/2}} \; \exp\left[E^-(\tau)\right] \; , \\  
\phi_2(\tau) &=& \sqrt{g^2+\tau^2} \; \exp\left[-\frac{\xi}{1-\R}Q(\tau)+\frac{i \gamma}{1-\R} \tau \right] \equiv \sqrt{g^2+\tau^2} \; \exp\left[E^+(\tau)\right] \; , 
\ENA
to leading order in $\xi^{-1}$. Here:
\EQA
\label{Definitions}
Q(x)&=&x^2/3+g^2x \; , \\   \label{Definitions2}
E^-(x)&=&-\xi /(1+\R) Q(x)- i \gamma / (1+\R) x \; , \\   \label{Definitions3}
E^+(x)&=&-\xi /(1-\R) Q(x)+ i \gamma / (1-\R) x \; .
\ENA
Using the method of variations of parameters, the full solution of Eq. (\ref{OneEquation}) with the initial conditions $\phi^{+}(\tau=\tau_0) = 0$ and $[\partial_\tau \phi^{+}] (\tau=\tau_0) = \tilde{f}_x(\tau=\tau_0) (g^2+\tau_0^2)/ \A $ is found as:
\EQ
\label{Weak+}
\phi^{+} \sim \int_{\tau_0}^\tau du \frac{(g^2+u^2) \tilde{f}_x(u)}{\A} \frac{\sqrt{g^2+\tau^2}}{\sqrt{g^2+u^2}} \exp\left[E^+(\tau)- E^+(u)\right] \; ,
\EN
where $E^+$ is defined in Eq. (\ref{Definitions3}). The $z$ component of the field can be obtained by integrating the $z$ component of Eq. (\ref{System2bis}) with the following result:
\EQA
\nonumber
\tilde{\psi}_z^{+} &\sim& \int_{\tau_0}^\tau du \frac{\tilde{f}_z(u)}{\A (1-\R)} \exp\left[E^+(\tau)- E^+(u)\right] \\  \label{Weakz}
&& - b  \int_{\tau_0}^\tau du \frac{\sqrt{g^2+u^2} \tilde{f}_x(u)}{\A (1-\R)} \exp\left[E^+(\tau)- E^+(u)\right] I^+(u,\tau) \; .
\ENA
Here we defined the integral $I^+$ as follows:
\EQ
\label{I++}
I^+(u,\tau) = \int_u^{\tau} \frac{dx}{(g^2+x^2)^{3/2}} \left\{1+\exp\left[-\frac{2 R \xi}{1-\R^2}\{Q(x)-Q(u)\} - \frac{2 i \gamma}{1-\R^2}(x-u) \right]\right\} \; .
\EN
The $y$-component can be obtained using the incompressibility condition.

\begin{figure}
\includegraphics[scale=0.7]{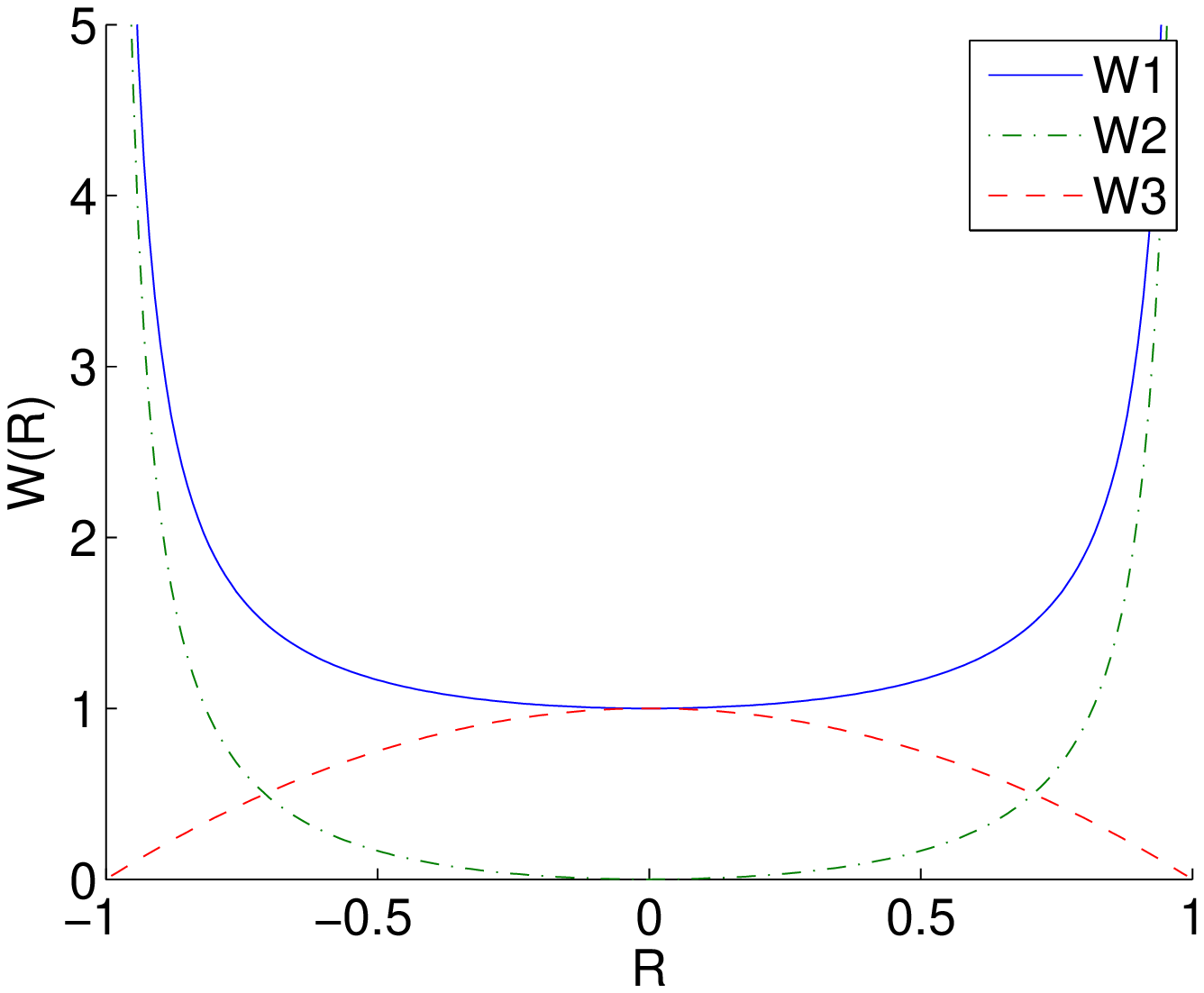}
\caption{\label{FigSR1} Dependence of $\W_1(R)$, $\W_2(R)$ and $\W_3(R)$ on the magnetic shear.}
\end{figure}

Using Eqs. (\ref{Weak+}-\ref{Weakz}), we can compute the  turbulent intensity and turbulent transport coefficients in a similar way as shown in the beginning of Appendix \ref{Integxi}. The results are as follows:
\EQA
\label{Weak2}
\langle u_x^2 \rangle &=& \frac{\tau_f}{(2\pi)^3 \A} \int d^3 k \frac{k_H^2}{k_y^2} E(k) I_{vx}({\bf k}) \; , \\  
\langle b_x^2 \rangle &=& \frac{\tau_f}{(2\pi)^3 \A} \int d^3 k \frac{k_H^2}{k_y^2} E(k) I_{bx}({\bf k}) \; , \\  
\langle u_z^2 \rangle &=& \frac{\tau_f}{(2\pi)^3 \A} \int d^3 k  E(k) I_{vz}({\bf k}) \; , \\  
\langle b_z^2 \rangle &=& \frac{\tau_f}{(2\pi)^3 \A} \int d^3 k  E(k) I_{bz}({\bf k}) \; , \\  
\mathcal{S} &=& \frac{\tau_f}{(2\pi)^3 \A} \int d^3 k \frac{E(k)}{1-\R^2} I_S({\bf k}) \; , \\  
\mathcal{E}_y &=& - \frac{\tau_f}{(2\pi)^3 \A} \int d^3 k  \frac{k_y}{k^2} \frac{H(k)}{1-\R^2} I_{\alpha}({\bf k})   \; , \\  \label{Weak2bis}
\mathcal{E}_z &=& \frac{\tau_f}{(2\pi)^3 \A} \int d^3 k  \frac{E(k)}{1-\R^2} I_\beta({\bf k})  \; , 
\ENA
where $k_H^2=k_y^2+k_z^2$ and the various integrals $I$'s are provided in Eq. (\ref{Integrals}-\ref{IntegralsBis}) of Appendix \ref{AppIntegrals}. In the limit $\xi \gg 1$, the approximate value of these integrals in  Eqs. (\ref{Weak2}-\ref{Weak2bis}) can be evaluated and then used for the computation of  the turbulent intensity and transport. The results are:
\EQA
\label{Weak3}
\langle u_x^2 \rangle &=&  \frac{\tau_f}{(2\pi)^3} \int d^3 k \frac{k_H^2 E(k)}{2 \nu k^4} \frac{1-\R^2/2}{1-\R^2}   \; , \\  
\langle b_x^2 \rangle &=&  \frac{\tau_f}{(2\pi)^3} \int d^3 k \frac{k_H^2 E(k)}{2 \nu k^4} \frac{\R^2/2}{1-\R^2}   \; , \\  
\langle u_z^2 \rangle &=&  \frac{\tau_f}{(2\pi)^3} \int d^3 k \frac{(k_x^2 + k_y^2) E(k)}{2 \nu k^4} \frac{1-\R^2/2}{1-\R^2} \; , \\  
\langle b_z^2 \rangle &=&  \frac{\tau_f}{(2\pi)^3} \int d^3 k \frac{(k_x^2 + k_y^2) E(k)}{2 \nu k^4} \frac{\R^2/2}{1-\R^2} \; , \\  
\nu_T &=& - \frac{\tau_f}{(2\pi)^3}  \int d^3 k \frac{E(k)(1-\R^2)}{4 \nu^2 k^8} [(k_x^2-k_H^2) k^2 + k_z^2 k_H^2] \; , \\  
\alpha_{yy} &=& - \frac{\tau_f}{2 (2\pi)^3} \int d^3 k  \frac{k_y^2 H(k)}{\nu^2 k^6}  \; , \\  \label{Weak3bis}
\beta &=& \frac{\tau_f}{(2\pi)^3} \int d^3 k \frac{k_z^2 E(k)(1-\R^2)}{\nu^2 k^6}   \; . 
\ENA 
Using the fact that the forcing is isotropic, Eqs. (\ref{Weak3}-\ref{Weak3bis}) can be simplified by integration over the angular variable $\theta$ and $\phi$ (after expressing the wave vector in spherical coordinates: $k_x = k \cos\theta$, $k_y = k \sin\theta\cos\phi$ and $k_z = k \sin\theta\sin\phi$). Eqs. (\ref{Weak3}-\ref{Weak3bis}) can then be recast as:  
\EQA
\label{Bull1}
\langle u_x^2 \rangle &=&  \frac{\tau_f}{(2\pi)^2} \int_0^{+\infty} d k \frac{2 E(k)}{3 \nu} \frac{1-\R^2/2}{1-\R^2}  \sim e_0 \W_1(\R) \; , \\  
\langle b_x^2 \rangle &=&  \frac{\tau_f}{(2\pi)^2} \int_0^{+\infty} d k \frac{2 E(k)}{3 \nu} \frac{\R^2/2}{1-\R^2}  \sim e_0 \W_2(\R) \; , \\  
\langle u_z^2 \rangle &=&  \frac{\tau_f}{(2\pi)^2} \int_0^{+\infty} d k \frac{2 E(k)}{3 \nu} \frac{1-\R^2/2}{1-\R^2} \sim e_0 \W_1(\R) \; , \\  
\langle b_z^2 \rangle &=&  \frac{\tau_f}{(2\pi)^2} \int_0^{+\infty} d k \frac{2 E(k)}{3 \nu} \frac{\R^2/2}{1-\R^2} \sim e_0 \W_2(\R) \; , \\  
\nu_T &=& \frac{\tau_f}{(2\pi)^2} \int_0^{+\infty} d k \frac{E(k)(1-\R^2)}{\nu^2 k^2} \frac{1}{30} \sim \frac{e_0}{3\nu k^2} \W_3(\R)  \; , \\  
\alpha_{yy} &=& - \frac{\tau_f}{(2\pi)^2} \int_0^{+\infty} d k \frac{H(k)}{3 \nu^2 k^2}  \sim \frac{h_0}{3 \nu k^2}  \; , \\  \label{Bull1bis}
\beta &=& \frac{\tau_f}{(2\pi)^2} \int_0^{+\infty} d k \frac{2 E(k)(1-\R^2)}{3\nu^2 k^2}  \sim \frac{2 e_0}{3 \nu k^2} \W_3(\R)  \; . 
\ENA 
In Eqs. (\ref{Bull1}-\ref{Bull1bis}), the functions $\W_n$, which are plotted on Fig. \ref{FigSR1}, characterize the dependence of turbulence intensity and transport on the magnetic shear. Figure \ref{FigSR1} shows that the magnetic shear tends to increase the turbulence intensity whereas it decreases the turbulent dissipation of both momentum and magnetic field ($\nu_T$ and $\beta$). Furthermore, the $\alpha$ effect is not affected by the magnetic shear and is the same as in the kinematic regime. The increase in the turbulence intensity can be understood in terms of instability of the homogeneous part of Eq. (\ref{OneEquation}). As can be seen in Eqs. (\ref{LargeTau}-\ref{LargeTau2}), the system tends to becomes unstable (the exponential decay being slower) when the magnetic shear is increased. These results are summarized in Table \ref{Summary}.

%\EQA
%\langle u_x^2 \rangle &=&  \frac{\tau_f}{(2\pi)^2} \int_0^{+\infty} d k \frac{2 E(k)}{3 \nu} \left[\frac{1-\R^2/2}{1-\R^2} + \frac{207 \xi_*^{-2}}{70} (1-\R^2+\R^4/2) \right]  \\  
%&\sim& e_0 \left[\frac{1-\R^2/2}{1-\R^2} + \frac{207 \xi_*^{-2}}{70} (1-\R^2+\R^4/2) \right] \; , \\  
%\langle u_z^2 \rangle &=&  \frac{\tau_f}{(2\pi)^2} \int_0^{+\infty} d k \frac{2 E(k)}{3 \nu} \left[\frac{1-\R^2/2}{1-\R^2} + \frac{2 \xi_*^{-2}}{35} (1-\R^2+\R^4/2) \right]  \\  
%&\sim& e_0 \left[\frac{1-\R^2/2}{1-\R^2} + \frac{2 \xi_*^{-2}}{35} (1-\R^2+\R^4/2) \right] \; , \\  
%\nu_T &=& \frac{\tau_f}{(2\pi)^2} \int_0^{+\infty} d k \frac{E(k)(1-\R^2)}{\nu^2 k^2} \left[\frac{4}{3} + \frac{568 \xi_*^{-2}}{45 (1-\R^2)^2}  \right]  \\  
%&\sim& \nu_0 (1-\R^2) \left[1 + \frac{142 \xi_*^{-2}}{15 (1-\R^2)^2}  \right]  \; , \\  
%\alpha_{yy} &=& \frac{\tau_f}{(2\pi)^2} \int_0^{+\infty} d k \frac{H(k)}{4 \nu^2} \left[\frac{2}{3} + \frac{11 \xi_*^{-2}}{35 (1-\R^2)^2}  \right]  \\  
%&\sim& \alpha_0 \left[1 + \frac{33 \xi_*^{-2}}{70 (1-\R^2)^2}  \right]   \; , \\  
%\beta &=& - \frac{\tau_f}{(2\pi)^2} \int_0^{+\infty} d k \frac{E(k)(1-\R^2)}{\nu^2 k^2} \left[\frac{2}{15} + \frac{31 \xi_*^{-2}}{210 (1-\R^2)^2}  \right] \; , \\  
%&\sim& \nu_0 (1-\R^2) \left[1 + \frac{465 \xi_*^{-2}}{420 (1-\R^2)^2}  \right]  \; . \\  
%\ENA 

\subsection{Strong Flow Shear: $\xi \ll 1$}
\label{StrongShear}
\begin{figure}
\includegraphics[scale=0.5]{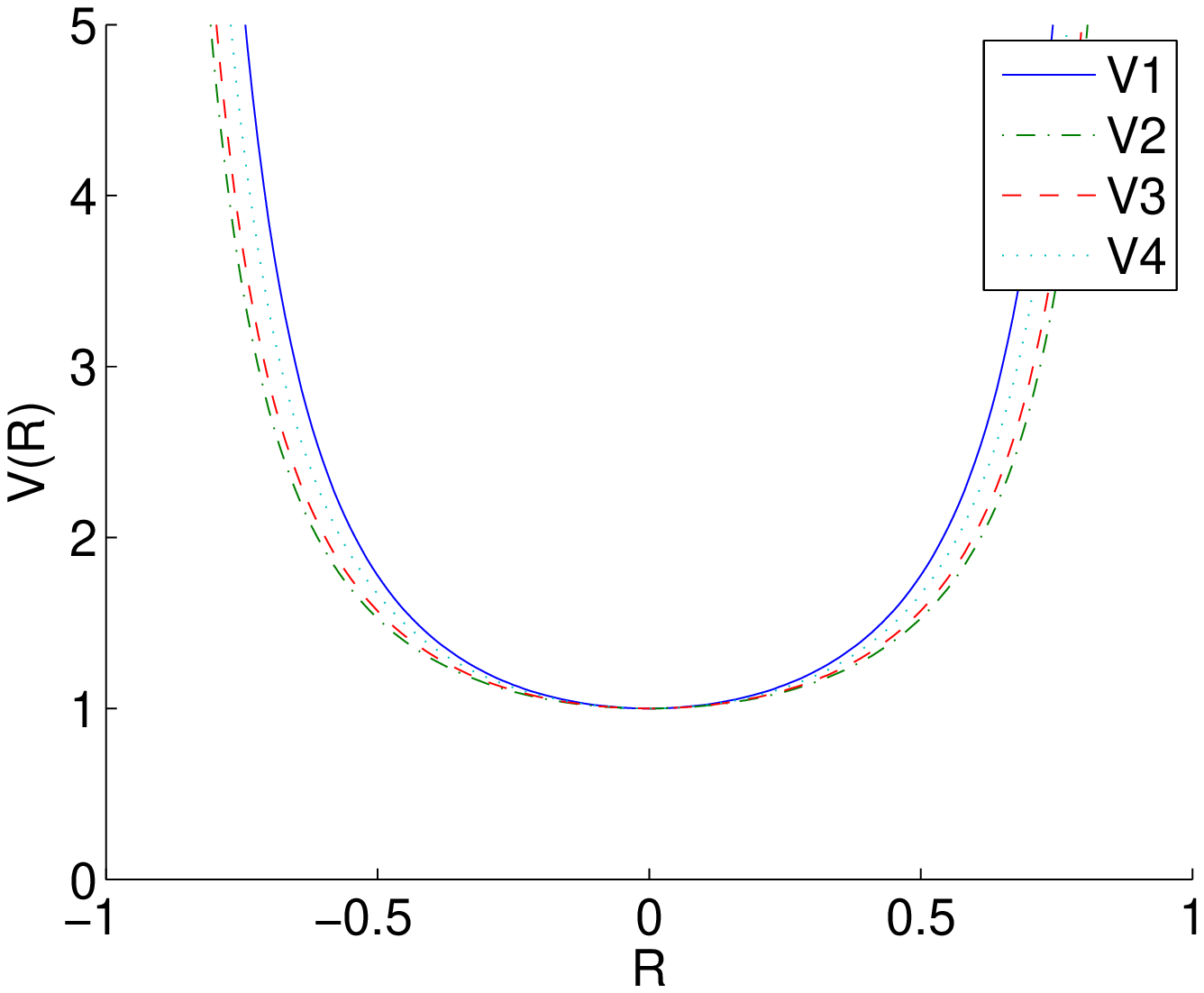}
\includegraphics[scale=0.5]{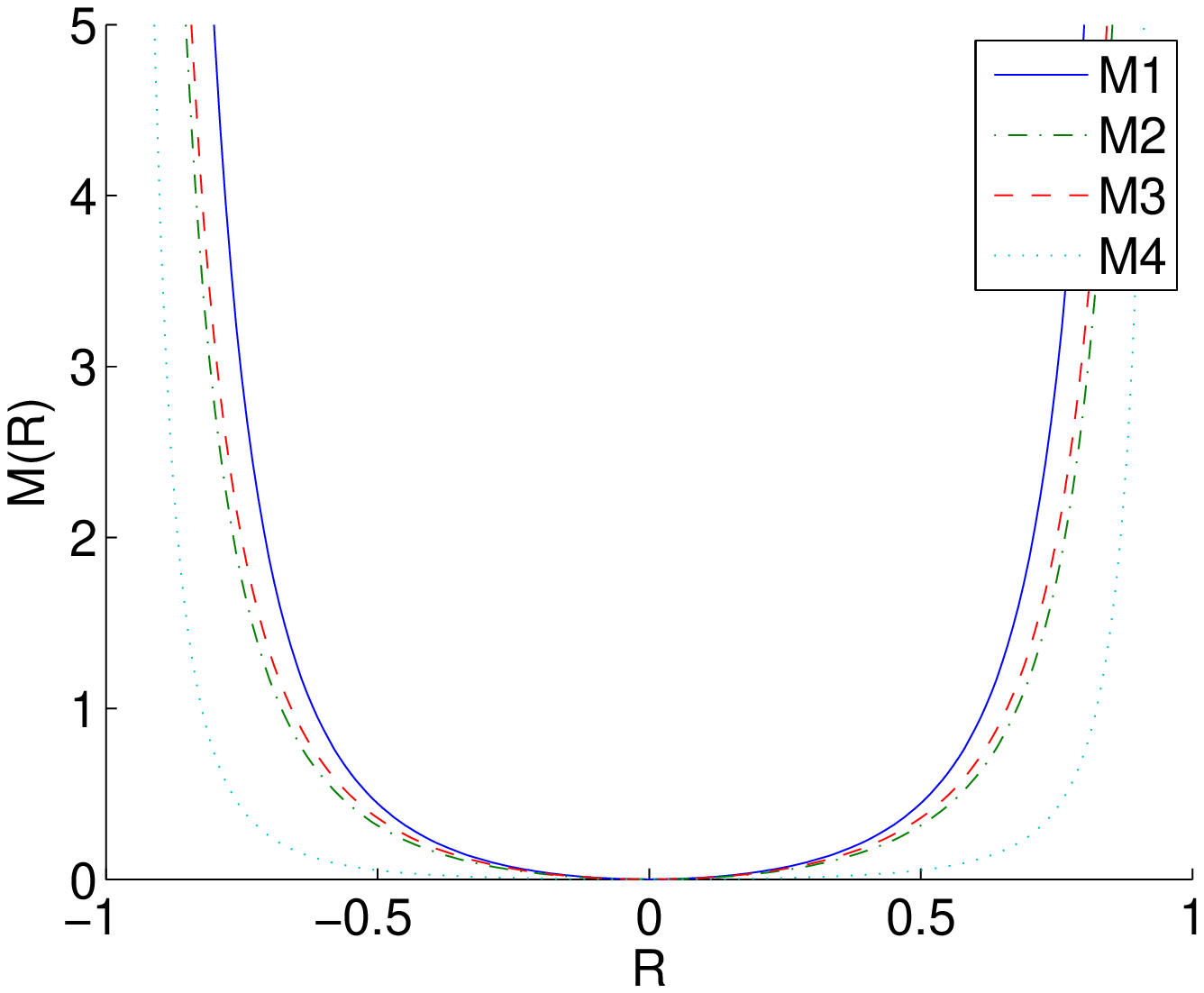}
\caption{\label{FigSR2} Functions characterizing the dependence on the magnetic shear of the velocity (left) and magnetic field (right).}
\end{figure}

The homogeneous part of Eq. (\ref{OneEquation}) has two independent solutions $\phi_1(\tau)$ and $\phi_2(\tau)$. Guided by the result of section \ref{LinearStab}, we rewrite these two solutions as: 
\EQA
\label{Bull3}
\phi_1(\tau) &=&  C_1(\tau) \exp\left[-\frac{\xi}{1+\R} Q(\tau) - \frac{i \gamma}{1+\R} \tau \right] \; , \\ 
\phi_2(\tau) &=& C_2(\tau) \exp\left[-\frac{\xi}{1-\R} Q(\tau) + \frac{i \gamma}{1-\R} \tau \right] \; .
\ENA
We can determine the asymptotic behavior of the unknown functions: $C_1(\tau) \sim \tau^{-2}$ and $C_2(\tau) \sim \tau$ and  for $\tau \gg 1$ (see appendix \ref{WKBLargeTau} for details). Using the method of variation of parameters, the general solution of Eq. (\ref{OneEquation}) is obtained as:
\EQA
\nonumber
\phi^+(\tau) &\sim& \int_{\tau_0}^\tau du \frac{(g^2+u^2) \tilde{f}_x(u)}{\A} \Bigl[ \frac{1}{u} \partial_u \left( \frac{u}{W(u)} \left\{C_2(\tau) C_1(u) e^{E^+(\tau)+E^-(u)} -C_2(u) C_1(\tau) e^{E^+(u)+E^-(\tau)} \right\} \right) \\ 
&&  - \frac{i \gamma + \xi (g^2+u^2)}{1+\R}  \frac{C_2(\tau) C_1(u) e^{E^+(\tau)+E^-(u)} -C_2(u) C_1(\tau) e^{E^+(u)+E^-(\tau)}}{W(u)} \Bigr] \; .
\ENA
Here $W(t)$ is the Wronskian of the two solutions of the homogeneous problem; $E_+$ and $E^-$ are defined in Eqs. (\ref{Definitions2}-\ref{Definitions3}). After a long algebra, we can compute all the components of the fields to leading order in $\xi$ with the following results:
\EQA
\phi^+(\tau) &\sim& \int_{\tau_0}^\tau du \frac{(g^2+u^2) \tilde{f}_x(u)}{\A} \left[C_3(u) C_2(\tau) e^{E^+(\tau)-E^+(u)} + C_4(u) C_1(\tau) e^{E^-(\tau)-E^-(u)}\right] \; , \\  \nonumber
\psi^+_z(\tau) &\sim& \int_{\tau_0}^\tau du \frac{\tilde{f}_z(u)}{(1-\R)\A}  e^{E^+(\tau)-E^+(u)}  \\  \label{psizlarge}
&& - \frac{b}{1-\R} \int_{\tau_0}^\tau du \frac{(g^2+u^2) \tilde{f}_x(u)}{\A} \left[e^{E^+(\tau)-E^+(u)} + e^{E^-(\tau)-E^-(u)}\right] C_5(u,\tau) \; .
\ENA
Here, note that the analytical form of the functions $C_3(u)$, $C_4(u)$ and $C_5(u,\tau)$ are unknown. However, their asymptotic behavior for large arguments ($\tau \rightarrow \infty$) can be computed from those of $C_1$ and $C_2$. In particular, in the large $\tau$ limit, the functions $C_5$ is independent of $\tau$. As shown in appendix \ref{Integxi}, the asymptotic dependence on $\tau$ is sufficient for computing the scaling of the turbulence intensity with the shear. By using similar techniques as in appendix \ref{Integxi}, the magnitude of the turbulent velocity and magnetic field can be found to leading order in $\xi$ as:
\EQA
\label{Bull2}
\langle u_x^2 \rangle &=& \frac{\tau_f}{(2\pi)^3 (1-\R^2)^2 \A} \int d^3 k \frac{k^2 k_H^2 E(k)}{k_y^4} J_{x}({\bf k}) \sim e_0 \xi \V_1(\R)  \; , \\  
\langle u_z^2 \rangle &=& \frac{\tau_f}{3 (2\pi)^3 \A} \int d^3 k E(k) \xi^{-1/3}  \left[J_{z1}({\bf k}) \V_2(\R) + J_{z2}({\bf k}) \V_3(\R) + J_{z3}({\bf k}) \V_4(\R) \right] \\ \nonumber 
&\sim&  e_0 \xi^{2/3} \left[\V_2(\R) + \V_3(\R) + \V_4(\R) \right]  \; , \\   
\langle b_x^2 \rangle &=& \frac{\tau_f \R^2}{(2\pi)^3 (1-\R^2)^2 \A} \int d^3 k \frac{k^2 k_H^2 E(k)}{k_y^4} J_{x}({\bf k}) \sim e_0 \xi \M_1(\R)  \; , \\  \label{Bull2bis}
\langle b_z^2 \rangle &=& \frac{\tau_f}{3 (2\pi)^3 \A} \int d^3 k E(k) \xi^{-1/3}  \left[J_{z1}({\bf k}) \M_2(\R) + J_{z2}({\bf k}) \M_3(\R) + J_{z3}({\bf k}) \M_4(\R) \right] \\  \nonumber 
&\sim&  e_0 \xi^{2/3} \left[\M_2(\R) + \M_3(\R) + \M_4(\R) \right]  \; .   
\ENA
Here, the $J$'s are convergent integrals which are independent of the velocity; $\V$'s and $\M$'s are defined as:
\EQA
\V_1(\R) &=& \frac{1}{(1-\R^2)^2} \; , \\  
\V_2(\R) &=& \frac{1}{4} \left[(1-\R)^{-5/3} + (1+\R)^{-5/3} + 2 (1-\R^2)^{-2/3} \right] \; , \\  
\V_3(\R) &=& \frac{1}{2} \left(\V_2(\R) + (1-\R^2)^{-5/3}\right) \; , \\  
\V_4(\R) &=& \frac{1}{4 (1-\R^2)^2} \left[(1-\R)^{1/3} + (1+\R)^{1/3} + 2 (1-\R^2)^{1/3} \right] \; , \\  
\M_1(\R) &=& \frac{\R^2}{(1-\R^2)^2} \; , \\  
\M_2(\R) &=& \frac{1}{4} \left[(1-\R)^{-5/3} + (1+\R)^{-5/3} - 2 (1-\R^2)^{-2/3} \right] \; , \\  
\M_3(\R) &=& \frac{1}{2} \left(\M_2(\R) + \R^2 (1-\R^2)^{-5/3} \right)\; , \\  
\M_4(\R) &=& \frac{1}{4 (1-\R^2)^2} \left[(1-\R)^{1/3} + (1+\R)^{1/3} - 2 (1-\R^2)^{1/3} \right] \; . \ENA
Eqs. (\ref{Bull2}-\ref{Bull2bis}) show that both the turbulent intensity and the magnetic field are reduced by strong flow shear $\A$. Furthermore, the quenching is anisotropic as the components in the direction of the shear ($\langle v_x^2 \rangle \sim \langle b_x^2 \rangle \sim \A^{-1}$) are much more reduced than the components in the perpendicular direction  ($\langle v_z^2 \rangle \sim \langle b_z^2 \rangle \sim \A^{-2/3}$) as $\A$ increases. Figure \ref{FigSR2} shows the dependence on the magnetic shear $\R$ of the velocity and magnetic field amplitude: they are increasing functions of the magnetic shear. This shows that the effect of the shear is to increase the turbulent intensity. In other words, the magnetic shear acts in the opposite way to the velocity shear, interfering with flow shear to weaken the quenching of turbulence by flow shear. 

In the limit $\xi \ll 1$, the Reynolds stress and the electromotive force can similarly be computed with the following turbulent transport coefficients:
\EQA
\label{Visco0}
\nu^T &\sim& \frac{\tau_f}{(2\pi)^3 \A^2} \int d^3 k E(k) \frac{ I_\nu({\bf k})}{1-\R^2}  \left\vert \ln\left(\frac{\xi }{1-\R^2}\right) \right\vert \sim  \frac{\xi^2 e_0}{(1-\R^2) \nu k^2}   \left\vert \ln\left(\frac{\xi }{1-\R^2}\right) \right\vert \; , \\  
\beta &\sim& \frac{\tau_f}{(2 \pi)^3 \A^2} \int d^3 k E(k) I_\beta({\bf k}) \frac{1}{1-\R^2} \frac{1}{\R} \ln\left(\frac{1+\R}{1-\R}\right) \sim \frac{e_0 \xi^2}{(1-\R^2) \nu k^2} \frac{1}{\R}  \ln\left(\frac{1+\R}{1-\R}\right) \; .
\ENA
Eq. (\ref{Visco0}) shows that the turbulent viscosity and diffusivity are reduced by a strong shear. Moreover, the reduction in the turbulent viscosity is weaker by a logarithmic factor than that in the turbulent diffusivity. Eq. (\ref{Bull3}) also shows that the effect of the magnetic shear is to increase the turbulent viscosity and diffusivity. It thus weakens the quenching effect of the shear flow. These results are again summarized in Table (\ref{Summary}).

%\subsection{Transport of particles}
%To leading order, we obtain the following for turbulent diffusivity:
%\EQA
%D^T_{xx} &=& \frac{\tau_f}{2 (1-\R^2)^2 (2\pi)^3 \A^2} \int d^3 k \frac{k^2 k_H^2 E(k)}{k_y^4} \T({\bf k})^2 \sim \frac{\xi^2}{(1-\R^2)^2} \nu_0 \; , \\  
%D^T_{zz}  &=& \frac{\tau_f}{3 (1-\R^2)^2 (2\pi)^3 \A^2} \int d^3 k E(k) \Gamma(2/3) \left( \frac{3}{2\xi}\right)^{2/3} \left[\frac{k_x^2+k_y^2}{k^2} + \frac{k_z^2 k^2}{k_y^4} \T({\bf k})^2 \right] \sim \frac{\xi^{4/3}}{(1-\R^2)^2} \nu_0 \; ,
%\ENA
%which shows that, similarly to the turbulence intensity, the transport of particle is reduced more severely in the direction of the shear. This results do not involve the magnetic field to leading order and consequently agree with the hydrodynamical case treated by \cite{Kim05}.

\section{Strong Magnetic field}
\label{StrongMagnetic}
For strong magnetic field ($\gamma = B_0 k_y / \A \gg 1$), a WKB analysis of Eq. (\ref{OneEquation}) gives:
\EQA
\label{ResultWKB}
\hat{\psi}^{+}_{x} &\sim& \frac{1}{\A (1-\R) \sqrt{g^2+\tau^2}} \int_{\tau_0}^\tau  \sqrt{g^2+u^2} \tilde{f}_x(u) \exp\left[E^+(\tau) - E^+(u)\right] du \; , \\  \nonumber
\hat{\psi}^{+}_{z} &\sim& \frac{1}{\A (1-\R)} \int_{\tau_0}^\tau \exp\left[E^+(\tau) - E^+(u) \right] \tilde{f}_z(u) du \\  \label{ResultWKB2}
&& - \frac{b}{\A (1-\R) } \int_{\tau_0}^\tau \exp\left[E^+(\tau) - E^+(u) \right] I(u,\tau)  \sqrt{g^2+u^2} \, \tilde{f}_x(u) \, du  \; ,
\ENA
to leading order in $\gamma^{-1}$. Here:
\EQ
I = \int_u^\tau \frac{du}{(g^2+u^2)^{3/2}} = \frac{1}{g^2} \left\{ \frac{\tau}{\sqrt{g^2+\tau^2}} - \frac{u}{\sqrt{g^2+u^2}}\right\} \; ,
\EN
and $E^+$ and $E^-$ are given in Eqs. (\ref{Definitions2}-\ref{Definitions3}).

The solution for the conjugate Elsasser variables, $\hat{\psi}^{-}_{x}$ and $\hat{\psi}^{-}_{z}$, can be obtained by changing the sign of $\gamma$ and $\R$ in Eqs. (\ref{ResultWKB}-\ref{ResultWKB2}). Using Eqs. (\ref{ResultWKB}-\ref{ResultWKB2}), we can compute the intensity of turbulence, with the following result to leading order in $\gamma^{-1}$:
\EQA
\label{WKBIntensity}
\langle u_x^2 \rangle &=&  \langle b_x^2 \rangle = \frac{\tau_f (1+\R^2)}{2 (1-\R^2)^2 (2\pi)^3 \A} \int d^3 k \frac{k_H^2 E(k)}{k_y^2} \T({\bf k}) \sim \frac{\xi (1+\R^2)}{(1-\R^2)^2} e_0 \; , \\  
\label{WKBIntensity2}
\langle u_z^2 \rangle &=& \langle b_z^2 \rangle = \frac{\tau_f}{3 (2\pi)^3 \A} \int d^3 k E(k) \S^0(R) \Gamma(1/3) \left( \frac{3}{2\xi}\right)^{1/3} \sim \xi^{2/3} \S^0(R) e_0 \; .
\ENA
Here, $k_H^2 = k_y^2 + k_z^2$ and $\T({\bf k}) = \pi \vert k_y \vert / 2 k_H - \arctan(k_x/k_H)$. The fact that the velocity and magnetic fields are in equipartition is due to a strong magnetic field which drives an Alfv\'enic turbulence. Note that the effect of the magnetic shear is included in the term:
\EQ
\S^0(R) = \frac{1}{4} \left[(1-\R)^{-5/3} + (1+\R)^{-5/3} \right] \; .
\EN
Eqs. (\ref{WKBIntensity}-\ref{WKBIntensity2}) show that the turbulence is unaffected by $B_0$ while severely quenched by flow shear. Furthermore, turbulence is less severely reduced in the direction perpendicular to the shear than in the direction of the shear. In particular, we see that, even in the limit of strong magnetic field, the turbulence intensity is reduced solely by the flow shear. This is because the magnetic field forces the turbulence to be more wave-like (due to Alfv\'en waves), thus increasing the memory time, without necessarily reducing its amplitude. Note that both  $\langle u_x^2 \rangle$ and $\langle u_z^2 \rangle$ increase with the magnetic shear. That is, the effect of the magnetic shear is again to weaken the quenching of the flow shear.

Similarly, the turbulent viscosity and the $\alpha$ effect are found as:
\EQA
\label{ViscoStrong}
\nu^T &=& \frac{\tau_f (1-\R^2)}{4 (2 \pi)^3} \int d^3 k \; E(k)  \frac{k_H^2 k_z^2 k_y^2 }{k^4} 
\frac{B_0^2}{[B_0^2 k_y^2+\nu k^2]^2} \; , \\  
\alpha_{yy} &=& - \frac{\tau_f}{2 (2\pi)^3} \int d^3 k \;  \frac{k_y^2 H(k)}{k^2(B_0^2 k_y^2 + \nu^2 k^2)} \; ,
\ENA
to leading order in $\gamma^{-1}$. The $\alpha$-effect is the same as in the case without magnetic shear showing that the magnetic shear has no effect on the $\alpha$-effect. In comparison, the turbulent viscosity $\nu^T$ is reduced as the magnetic shear increases. Furthermore, we find that the turbulent diffusivity $\beta$ vanishes to leading order. Non-trivial diffusivity will be found only at the higher order in $\gamma^{-1}$. This means that magnetic fields hardly diffuse when they are too strong, which is in agreement with numerical simulations \cite{Newton08}.

\section{Conclusion}
\label{Discussion}
To understand the properties of astrophysical and geophysical magnetic fields, we investigated the turbulent transport of non-uniform magnetic field in the presence of flow shear. The linear analysis reveals that magnetic shear stronger than flow shear leads to instability. We considered the two limits of strong and weak magnetic field. In the weak magnetic field limit, the magnitude of the magnetic shear is limited only by the stability condition that the magnetic shear is to be weaker than the flow shear.

In the case of a weak magnetic field, the flow shear is shown to reduce both the turbulence intensity and turbulent transport due to shear stabilization in agreement with our previous results \cite{Kim05,2Shears}. In particular, turbulent viscosity and magnetic diffusivity ($\beta$ effect) are strongly suppressed as $\A^{-2}$ for strong flow shear $\A$. When magnetic shear is incorporated the turbulence intensity is increased. This is due to the fact that magnetic shear tends to make the system unstable as shown by our stability analysis. In contrast, the turbulent transport (both turbulent viscosity and diffusivity) is reduced by magnetic shear if the flow shear is weaker than diffusion rate while it is increased if the shear is stronger. In all the cases, we found that, for strong shear, the magnetic shear opposes the effect of flow shear, thereby compensating the quenching of turbulence by shear stabilization. These results are summarized in Table \ref{Summary}.

\renewcommand{\arraystretch}{1.8}
\begin{table}
\begin{center}
\begin{tabular}{|c|c|c|}
\hline
& \;  Weak flow shear $\xi= \nu k_y^2 / \A \gg 1 \; $ & \; Strong flow shear $\xi \ll 1 \; $ \\ \hline
$ \; \langle v_x^2 \rangle \;$ & $f_+(\R)$ & $\xi f_+(\R)$ \\
$ \; \langle v_y^2 \rangle \sim \langle v_z^2 \rangle \; $ & $f_+(\R)$  & $\xi^{2/3} f_+(\R)$ \\ \hline
$ \; \langle b_x^2 \rangle \;$ & $f_+(\R)$ & $\xi f_+(\R)$ \\
$ \; \langle b_y^2 \rangle \sim \langle b_z^2 \rangle \; $ & $f_+(\R)$  & $\xi^{2/3} f_+(\R)$ \\ \hline
$  \; \nu_T  \; $ & $f_-(\R)$ & $\xi^2 f_+(\R)$ \\ \hline
$ \; \beta \; $ & $f_-(\R)$ & $\xi^2 f_+(\R)$ \\  \hline
\end{tabular}
\end{center}
\caption{\label{Summary} Summary of our results obtained in the weak magnetic field limit. $f_+(R)$ and $f_-(R)$ represent a function which increases or decreases with magnetic shear $\R=B_1/\A$, respectively.}
\end{table}

For strong magnetic field, we found that the turbulent intensity is not quenched by strong magnetic field. In comparison, the turbulent viscosity and $\alpha$ effect are reduced by strong magnetic field with scalings $B_*^{-1}$ and $B_*^{-2}$, respectively. Furthermore, we found that the turbulent diffusivity of magnetic field (the $\beta$ effect) vanishes for strong magnetic field to leading order, indicating that strong magnetic fields hardly diffuse. To recapitulate, the $\beta$ effect is quenched by magnetic field for a large constant magnetic field whereas in the weak magnetic field limit, it can be reduced by strong flow shear.

It will be interesting to extend our theory to incorporate the effects of rotation which will consistently give rise to $\alpha$ effect and non-diffusive momentum transport ($\Lambda$ effect) due to shear-induced anisotropy \cite{RotShearAA}. How the $\Lambda$ effect, $\alpha$ effect, turbulent viscosity and particle transport are affected by rotation, magnetic field and shear would be of great interest with important practical implications. These issues will be addressed in future publications.

\begin{acknowledgments}
This work was supported by U.K. STFC Grant No. ST/F501796/1.
\end{acknowledgments}

\begin{appendix}
\section{Kinetic energy and $\alpha$ effect in the absence of shear}
\label{AppendixA}
In the case without large-scale shear flow, the linearized equation for the fluctuating velocity can be written as:
\EQ
\partial_t \uu({\bf x},t) = - {\bf \nabla} p({\bf x},t)  + \nu \Delta \uu({\bf x},t) + {\bf f}({\bf x},t) \; .
\EN
In the case where the forcing is incompressible, the pressure vanishes ($p=0$) and the solution of this equation can easily be obtained in Fourier space as:
\EQ
\label{SolSansShear}
\tilde{\uu}({\bf k},t) = \int_0^t du \; {\bf \tilde{f}}({\bf k},u) \exp[-\nu k^2 (t-u)] \; .
\EN
Using Eq. (\ref{SolSansShear}), the turbulent intensity can be computed as follows:
\EQA
\nonumber
\langle u^2 \rangle &=& \frac{1}{(2 \pi)^6} \int d^3 k_1 d^3 k_2 \; e^{i({\bf k_1}+{\bf k_2})\cdot {\bf x}} \int_0^t du_1 \int_0^t du_2 \; e^{-\nu \{k_1^2 (t-u_1)+ k_2^2 (t-u_2)\}} \, \langle {\bf \tilde{f}}({\bf k_1},u_1) \cdot {\bf \tilde{f}}({\bf k_2},u_2) \rangle \\ \label{SolSansShear2}
&=& \frac{\tau_f}{(2 \pi)^3} \int d^3 k_1 \int_0^t du_1 \; e^{- 2 \nu k_1^2 (t-u_1)} \,  2 E(k_1) \\ \nonumber
&=& \frac{\tau_f}{(2 \pi)^3} \int d^3 k_1  \frac{E(k_1)}{\nu k_1^2} \left(1-e^{- 2 \nu k_1^2 t}\right) 
\\ \nonumber
&\sim& \frac{\tau_f}{(2 \pi)^3} \int d^3 k_1  \frac{E(k_1)}{\nu k_1^2} \\ \nonumber
&=& \frac{2 \tau_f}{(2 \pi)^2} \int_0^{+\infty} d k_1 \;  \frac{E(k_1)}{\nu} 
\ENA
The second line in Eq. (\ref{SolSansShear2}) is obtained by using the correlation function of the forcing given by Eq. (\ref{Forcing}). The fourth line is obtained by taking the long-time limit ($t \rightarrow \infty$) while the last line is obtained by integrating over the angular variables.

Similarly the kinetic helicity can be obtained as:
\EQA
\nonumber
\langle {\bf u} \cdot ({\bf \nabla \times u}) \rangle &=& \frac{1}{(2 \pi)^6} \int d^3 k_1 d^3 k_2 \; e^{i({\bf k_1}+{\bf k_2})\cdot {\bf x}} \int_0^t du_1 \int_0^t du_2 \; e^{-\nu \{k_1^2 (t-u_1)+ k_2^2 (t-u_2)\}} \times \\ 
&& \langle \epsilon_{lmp} \; (i k_{2m}) \tilde{f}_l({\bf k_1},u_1) \tilde{f}_p({\bf k_2},u_2) \rangle \\ \nonumber
&=& \frac{\tau_f}{(2 \pi)^3} \int d^3 k_1  \;  \int_0^t du_1 \; e^{-2 \nu k_1^2 (t-u_1)} \epsilon_{lmp} (- i k_{1m})  \kappa_{lp}(-{\bf k_1}) \\ \nonumber
&=& \frac{\tau_f}{(2 \pi)^3} \int d^3 k_1 \int_0^t du_1 \; e^{- 2 \nu k_1^2 (t-u_1)} \,  2 H(k_1) \\ \nonumber
&=& \frac{\tau_f}{(2 \pi)^3} \int d^3 k_1  \frac{H(k_1)}{\nu k_1^2} \left(1-e^{- 2 \nu k_1^2 t}\right) 
\\ \nonumber
&\sim& \frac{\tau_f}{(2 \pi)^3} \int d^3 k_1  \frac{H(k_1)}{\nu k_1^2} \\ \nonumber
&=& \frac{2 \tau_f}{(2 \pi)^2} \int_0^{+\infty} d k_1 \;  \frac{H(k_1)}{\nu} 
\ENA

\section{WKB analysis in the long time limit}
\label{WKBLargeTau}
To study the behavior of Eq. (\ref{OneEquation}) for large time, we introduce a small parameter $\epsilon$ and write $\tau = x / \epsilon$. In terms of the new variable $x$, the homogeneous part of Eq. (\ref{OneEquation}) can be rewritten as:
\EQA
\label{WKBHomogeneous}
&& \qquad \qquad \epsilon^2 \partial_x^2 \phi^{+}(x) + \epsilon \left[-\frac{\epsilon}{x}  + \frac{2 \xi}{1-\R^2} \left(g^2+\frac{x^2}{\epsilon^2}\right) - \frac{2 i \gamma \R}{1-\R^2} \right] \partial_x \phi^{+}(x) \\ \nonumber
&+& \left[\frac{i \gamma \epsilon^2}{(g^2 \epsilon^2+x^2)} \left(\frac{\gamma \epsilon}{(1-\R)x} -\frac{x}{\epsilon (1+\R)}  \right)  + \frac{\xi}{1-\R} \left( \frac{x^2- \epsilon^2 g^2}{\epsilon x} + \frac{2 \R x}{ \epsilon (1+\R)}\right) + \frac{\xi^2 \left[g^2 +\left(\frac{x}{\epsilon}\right)^2\right]^2 +\gamma^2}{1-\R^2} \right] \phi^{+}(x) = 0 \; .
\ENA
Using the following WKB ansatz:
\EQ
\label{WKBansatz}
\phi^+(x) = \exp\left[\frac{1}{\epsilon^3}\left(S_0(x) + \epsilon S_1(x) + \dots \right) \right] \; ,
\EN
and solving order by order (in $\epsilon$), we can obtain the two solutions of the homogeneous equation (\ref{WKBHomogeneous}) with the following values of $S_0$, $S_1$, $S_2$ and $S_3$:
\renewcommand{\arraystretch}{2.8}
\begin{center}
\begin{tabular}{|c|c|c|}
\hline
 & Solution 1 & Solution 2 \\ \hline
$\qquad S_0(x) \qquad$ & $\qquad - \frac{\xi x^3}{3 (1+\R)} \qquad$ & $\qquad - \frac{\xi x^3}{3 (1-\R)} \qquad$ \\ \hline
$\qquad S_1(x) \qquad$ & $0$ & $0$  \\ \hline
$\qquad S_2(x) \qquad$ & $-\frac{g^2+ig}{1+\R} x$ & $-\frac{g^2-ig}{1-\R} x$ \\ \hline
$\qquad S_3(x) \qquad$ & $-2 \ln x$ & $\ln x$ \\ \hline
\end{tabular}
\end{center}
Plugging this result into Eq. (\ref{WKBansatz}) and changing back to the original variable $\tau=x / \epsilon$, we obtain the two following approximate solutions to the homogeneous equation:
\EQA
\phi_1(\tau) &\sim& \frac{1}{\tau^2}
 \exp\left[-\frac{\xi}{1+\R} Q(\tau) - \frac{i \gamma}{1+\R} \tau  + \frac{1}{\tau} l_1(\tau) \right] \; , \\ 
\phi_2(\tau) &\sim& \tau \exp\left[-\frac{\xi}{1-\R} Q(\tau) + \frac{i \gamma}{1-\R} \tau + \frac{1}{\tau} l_2(\tau) \right]  \; .
\ENA
Here, $Q(x) = g^2 x + x^3 / 3$, $l_1$ and $l_2$ are two functions which converge in the large $\tau$ limit.

\section{Integrals in the weak shear limit}
\label{AppIntegrals}
In the computation of the turbulent intensity and transport coefficients, we obtained the following integrals in Eq. (\ref{Weak2}-\ref{Weak2bis}):
\EQA
\label{Integrals}
I_{vx} &=& \int_a^{+\infty} \frac{d\tau}{4 (g^2+\tau^2)} \left\{\frac{e^{2E_0^+(a,\tau)}}{(1-\R)^2} + \frac{e^{E_0^-(a,\tau)}}{(1+\R)^2} + \frac{2 e^{2E_0(a,\tau)}}{1-\R^2} \cos\left[\frac{2 \gamma}{1-\R^2} (\tau-a)\right]\right\}  \; , \\  
I_{bx} &=& \int_a^{+\infty} \frac{d\tau}{4 (g^2+\tau^2)} \left\{\frac{e^{2E_0^+(a,\tau)}}{(1-\R)^2} + \frac{e^{E_0^-(a,\tau)}}{(1+\R)^2} - \frac{2 e^{2E_0(a,\tau)}}{1-\R^2} \cos\left[\frac{2 \gamma}{1-\R^2} (\tau-a)\right]\right\}  \; , \\  
I_{vz} &=& \int_a^{+\infty} d\tau \frac{1+a^2}{4(g^2+a^2)} \left\{\frac{e^{E_0^-(a,\tau)}}{(1-\R)^2} + \frac{e^{2E_0^+(a,\tau)}}{(1+\R)^2} + \frac{2 e^{2E_0(a,\tau)}}{1-\R^2} \cos\left[\frac{2 \gamma}{1-\R^2} (\tau-a)\right]\right\}   \\  \nonumber
&& + \frac{a b^2}{\sqrt{g^2+a^2}} \left\{\frac{e^{E_0^+(a,\tau)}}{(1-\R)^2} I_c^+ + \frac{e^{2E_0^-(a,\tau)}}{(1+\R)^2} I_c^- + \frac{e^{2E_0(a,\tau)}}{1-\R^2} \text{Rl} \left( e^{2 i \gamma (\tau-a) / (1-\R^2)} [I^+(a,\tau)+ I^-(a,\tau)] \right) \right\} \\  \nonumber
&& + \frac{b^2 g^2}{g^2+a^2} \left\{\frac{e^{E_0^+(a,\tau)}}{(1-\R)^2} \vert I_0^+(a,\tau) \vert^2 + \frac{e^{2E_0^-(a,\tau)}}{(1+\R)^2} \vert I_0^-(a,\tau) \vert^2 + \frac{e^{2E_0(a,\tau)}}{1-\R^2} \text{Rl} \left( e^{2 i \gamma (\tau-a) / (1-\R^2)} [I^+(a,\tau) I^-(a,\tau)^*] \right) \right\} \; , \\  
I_{bz} &=& \int_a^{+\infty} d\tau \frac{1+a^2}{4(g^2+a^2)} \left\{\frac{e^{E_0^-(a,\tau)}}{(1-\R)^2} + \frac{e^{2E_0^+(a,\tau)}}{(1+\R)^2} - \frac{2 e^{2E_0(a,\tau)}}{1-\R^2} \cos\left[\frac{2 \gamma}{1-\R^2} (\tau-a)\right]\right\}   \\  \nonumber
&& + \frac{a b^2}{\sqrt{g^2+a^2}} \left\{\frac{e^{E_0^+(a,\tau)}}{(1-\R)^2} I_c^+ + \frac{e^{2E_0^-(a,\tau)}}{(1+\R)^2} I_c^- - \frac{e^{2E_0(a,\tau)}}{1-\R^2} \text{Rl} \left( e^{2 i \gamma (\tau-a) / (1-\R^2)} [I^+(a,\tau)+ I^-(a,\tau)] \right) \right\} \\  \nonumber
&& + \frac{b^2 g^2}{g^2+a^2} \left\{\frac{e^{E_0^+(a,\tau)}}{(1-\R)^2} \vert I_0^+(a,\tau) \vert^2 + \frac{e^{2E_0^-(a,\tau)}}{(1+\R)^2} \vert I_0^-(a,\tau) \vert^2 - \frac{e^{2E_0(a,\tau)}}{1-\R^2} \text{Rl} \left( e^{2 i \gamma (\tau-a) / (1-\R^2)} [I^+(a,\tau) I^-(a,\tau)^*] \right) \right\} \; , \\  
I_{S} &=& \int_a^{+\infty} \frac{d\tau}{\sqrt{g^2+\tau^2}} e^{2E_0(a,\tau)} \Bigl\{ \cos\left[\frac{2 \gamma}{1-\R^2} (\tau-a)\right] \left(- \frac{\tau (g^2+a^2)}{\sqrt{g^2+\tau^2}} + \frac{a b^2}{\sqrt{g^2+a^2}} \right)  \\  \nonumber
&& + \frac{b^2 (g^2+a^2)}{2}  \text{Rl} \left( e^{2 i \gamma (\tau-a) / (1-\R^2)} [I^+(a,\tau) + I^-(a,\tau)] \right) \Bigr\}  \; , \\  
I_{\alpha} &=& \int_a^{+\infty} d\tau \frac{\sqrt{g^2+a^2}}{\sqrt{g^2+\tau^2}} e^{2E_0(a,\tau)} k_y \sin\left[\frac{2 \gamma}{1-\R^2} (\tau-a)\right]  \; , \\  \label{IntegralsBis}
I_{\beta} &=& \int_a^{+\infty} \frac{d\tau}{\sqrt{g^2+\tau^2}} e^{2E_0(a,\tau)} \frac{b^2 (g^2+a^2)}{2}  \text{Rl} \left( e^{2 i \gamma (\tau-a) / (1-\R^2)} [I^+(a,\tau) - I^-(a,\tau)] \right)   \; . 
\ENA
Here, $\text{Rl}$ stands for the real part; $I^+$ is defined in Eq. (\ref{I++}) and $I^-$ is obtained from $I^+$ by changing the signs of $\R$ and $\gamma$; we defined $a=k_x/ k_y$ and the following functions:
\EQA
\label{IntegralsI}
E_0^+(a,\tau)&=&-\frac{\xi}{1-\R} \left[Q(\tau)-Q(a)\right] \; , \\  
E_0^-(a,\tau)&=&-\frac{\xi}{1+\R} \left[Q(\tau)-Q(a)\right] \; , \\  
E_0(a,\tau)&=&-\frac{\xi}{1-\R^2} \left[Q(\tau)-Q(a)\right] \; , \\  
I_0^+(a,\tau) &=& \int_a^{\tau} \frac{dx}{(g^2+x^2)^{3/2}} \left\{1+\exp\left[-\frac{2 R \xi}{1-\R^2}\{Q(x)-Q(a)\} \right]\right\} \; , \\  
I_c^+(a,\tau) &=& \int_a^{\tau} \frac{dx}{(g^2+x^2)^{3/2}} \left\{1+\exp\left[-\frac{2 R \xi}{1-\R^2}\{Q(x)-Q(a)\} \right] \cos\left[\frac{2 \gamma}{1-\R^2}(x-a)\right] \right\} \; , \\  \label{IntegralsIbis}
I_s^+(a,\tau) &=& \int_a^{\tau} \frac{dx}{(g^2+x^2)^{3/2}} \left\{1+\exp\left[-\frac{2 R \xi}{1-\R^2}\{Q(x)-Q(a)\} \right] \sin\left[\frac{2 \gamma}{1-\R^2}(x-a)\right] \right\} \; .
\ENA
The integrals $I_0^-(a,\tau)$, $I_c^-(a,\tau)$ and $I_s^-(a,\tau) $ are obtained by changing the signs of $\R$ and $\gamma$ in Eqs. (\ref{IntegralsI}-\ref{IntegralsIbis}) .

\section{Asymptotic expansion}
Expanding all the integrals in Eq. (\ref{Integrals}) in powers of $\xi^{-1}$, we obtain the following two  leading orders (by keeping only the terms which are even in all wave-numbers as the terms with an odd number of wave-numbers would vanish upon  angular integration):
\EQA
I_{vx} &=& \frac{1}{8 \xi (g^2+a^2)^2} \left(\frac{1}{(1-\R)} + \frac{1}{(1+\R)} + 2 \right) = \frac{1}{2 \xi (g^2+a^2)^2} \frac{1-\R^2/2}{1-\R^2} \; , \\  
I_{bx} &=& \frac{1}{8 \xi (g^2+a^2)^2} \left(\frac{1}{(1-\R)} + \frac{1}{(1+\R)} - 2 \right) = \frac{1}{2 \xi (g^2+a^2)^2} \frac{\R^2/2}{1-\R^2} \; , \\  
I_{vz} &=& \frac{1+a^2}{8 \xi (g^2+a^2)^2} \left(\frac{1}{(1-\R)} + \frac{1}{(1+\R)} + 2 \right) = \frac{1+a^2}{2 \xi (g^2+a^2)^2} \frac{1-\R^2/2}{1-\R^2} \; , \\  
I_{bz} &=& \frac{1+a^2}{8 \xi (g^2+a^2)^2} \left(\frac{1}{(1-\R)} + \frac{1}{(1+\R)} - 2 \right) = \frac{1+a^2}{2 \xi (g^2+a^2)^2} \frac{\R^2/2}{1-\R^2} \; , \\  
I_{S} &=& \frac{a^2(g^2+a^2)-g^2}{4 \xi^2 (g^2+a^2)^4} (1-\R^2)^2 \; ,  \\  
I_{\alpha} &=& \frac{\gamma (1-\R^2)}{2 \xi^2 (g^2+a^2)^2}   \; , \\  
I_{\beta} &=& \frac{b^2 \R}{\xi^2 (g^2+a^2)^3} (1-\R^2)^2    \; . 
\ENA

\section{Integrals in the strong flow shear limit}
\label{Integxi}
As an example of how to compute the correlation functions, we show the main steps to obtain $\langle X^2 \rangle$. The Fourier transform $\tilde{X}$ of $X$ is given by:
\EQ
\tilde{X} = \int_{\tau_0}^\tau du \;  G(u) F(\tau) \exp\left[- \xi \left\{Q(\tau) - Q(u)\right\}\right] f_i(u) \; ,
\EN
where $Q$ is defined in Eq. (\ref{Definitions}). The correlation of the variable $X$ can then be computed as \citep[see][for details]{Kim05}:
\EQ
\langle X^2 \rangle = \frac{\tau_f}{(2 \pi)^3  \A} \int d^3 k \; K_{ii}({\bf k}) \, G(a)^2 \int_a^{+\infty} F^2(\tau) \exp\left[- \xi \left\{Q(\tau) - Q(a)\right\}\right] d\tau 
\EN
where $a=k_x/k_y$.

For the computation of the correlation in the strong shear limit ($\xi \ll 1$), we have to compute $\tau$-integrals of the form:
\EQ
\label{tauInteg}
K = \int_a^{+\infty} F^2(\tau) \exp\left[- \xi \left\{Q(\tau) - Q(a)\right\}\right]  d\tau \;  .
\EN  
Here $F^2$ has the scaling $F^2(\tau) \sim \tau^{\theta}$ in the large $\tau$ limit.  When $\theta < -1$, $K$ exists for $\xi=0$ thus, the $\xi \ll 1$ limit can be easily obtained by putting $\xi =0$ in Eq. (\ref{tauInteg}). This is however not the case when $\theta > -1$ as the integral diverges as $\xi \rightarrow 0$. In that case, by making the substitution $y = 2 \xi \tau^3 / 3$, the integral (\ref{tauInteg}) can be computed  in the $\xi \ll 1$ limit as:
\EQA
\label{tauInteg2}
K &=& \int_{2 \xi a^3 /3}^{+\infty} F^2\left[\left(\frac{3y}{2\xi}\right)^{1/3}\right] \exp\left[- y +  \left(\frac{3y \xi^2}{2}\right)^{1/3} \right] \left(\frac{3 y}{2 \xi}\right)^{-2/3} \frac{dy}{2\xi} \\ \nonumber
&\sim& \int_{0}^{+\infty} \exp[- y] \left(\frac{3 y}{2 \xi}\right)^{(\theta-2)/3} \frac{dy}{2\xi} \\ \nonumber
&\sim& \frac{1}{3} \left(\frac{3}{2 \xi}\right)^{(\theta+1)/3} \int_{0}^{+\infty} \exp[- y]  y^{(\theta-2)/3} dy \\ \nonumber
&\sim& \frac{1}{3} \left(\frac{3}{2 \xi}\right)^{(\theta+1)/3} \Gamma\left(\frac{\theta+1}{3}\right) \; ,
\ENA  
where $\Gamma$ is the Gamma function. In summary, if the integrand has a scaling $\tau^{\theta}$ (excluding the exponential factor) in the large $\tau$ limit, the correlation function scales as $\xi^{-(\theta+1)/3}$ in the strong shear limit ($\xi \ll 1$). For instance, Eq. (\ref{psizlarge}) shows that in the large $\tau$ limit, the integrand does not depend on $\tau$ ($\theta=0$). Therefore, for the computation of $\langle u_z^2 \rangle$, the $\tau$ integral has the scaling $\xi^{-1/3}$ with the shear. 

\end{appendix}

\bibliographystyle{apsrev}
\bibliography{../../../Biblio/Bib_sun,../../../Biblio/Bib_maths,../../../Biblio/Bib_dynamo,../../../Biblio/Bib_shear,Bib_3DMHD}
%\bibliography{Bib_3DMHD}

\end{document}